\documentstyle[psfig,12pt]{article}

\def\ell{l}

\def\bqll{$b \rightarrow q \ell^+ \ell^- $ }

\def\bxqll{$B \rightarrow X_q \ell^+ \ell^- $ }
\def\bxsll{$B \rightarrow X_s \ell^+ \ell^- $ }
\def\bxdll{$B \rightarrow X_d \ell^+ \ell^- $}

\def\BXu{$B \rightarrow X_u l \bar{\nu}$}
\def\absvcb{\left| V_{cb} \right|}

\def\s{\hat{s}}
\def\u{\hat{u}}

\def\c910{C_9^{\mbox{eff}} \mp C_{10}}

\def\mc{\hat{m}_c}

\def\mq{\hat{m}_q}
\def\lo{\hat{\lambda}_1}
\def\lt{\hat{\lambda}_2}
\def\loo{\lambda_1}
\def\lto{\lambda_2}

\def\Vtd{V_{td}}
\def\Vts{V_{ts}}
\def\Vuq{V_{uq}}
\def\Vub{V_{ub}}
\def\Vtq{V_{tq}}
\def\Vtb{V_{tb}}
\def\Vcq{V_{cq}}
\def\Vcb{V_{cb}}
\def\dBs{$ {d{\cal{B}} \over ds} [\mbox{\bxsll}] $}
\def\dBd{$ {d{\cal{B}} \over ds} [\mbox{\bxdll}] $}
\def\dBq{$ {d{\cal{B}} \over ds} [\mbox{\bxqll}] $}
\title{
$\mbox{\bxqll} (q=d,s)$ and 
Determination of $\left|{V_{td} \over V_{ts}}\right|$  
} 

\author{\vspace*{1em}\\
    {\bf C. S. Kim}\thanks{E-mail address: 
       kim@cskim.yonsei.ac.kr,~ cskim@kekvax.kek.jp} \\
  Department of Physics, Yonsei University \\
Seoul 120-749, Korea\\ 
    {\bf T. Morozumi}\thanks{E-mail address:
  morozumi@theo.phy.sci.hiroshima-u.ac.jp} \\
Department of Physics, Hiroshima University\\
        1-3-1 Kagamiyama, Higashi Hiroshima, Japan 739\\
    {\bf A. I. Sanda}\thanks{E-mail address: 
      sanda@eken.phys.nagoya-u.ac.jp}  \\
Department of Physics, Nagoya University\\
       Chikusa-ku, Nagoya, Japan 464-01 \\
             }
\date{}
\begin{document}
\setlength{\baselineskip}{20pt}
\maketitle
\begin{picture}(0,0)
       \put(325,350){YUMS-97-020}
       \put(325,335){SNUTP 97-094}
       \put(325,320){HUPD-9710}
       \put(325,305){DPN-97-34}
\end{picture}

\setlength{\baselineskip}{20pt}

\begin{abstract}

We propose a new method to extract $ |\Vtd| \over |\Vts|$
from  the ratio of the decay distributions 
${dR \over ds} \equiv \mbox{\dBd} / \mbox{\dBs}$.
This ratio depends only on the KM ratio 
${|\Vtd| \over |\Vts|}$
with   $15\%$ theoretical uncertainties, 
if dilepton invariant mass-squared $s$ is away
from the peaks of the  possible resonance states,  
$J/\psi$, $\psi'$, and {\it etc.}
We also give a detailed {\it analytical and numerical} 
analysis on $\mbox{\dBq}(q=d,s)$ and ${dR \over ds}$.
\end{abstract}
\thispagestyle{empty}
\newpage

\section{\bf Introduction}

The determination of the elements of KM matrix is one of
the most important issues in the quark flavor physics.
Moreover, the element $V_{td}$ (or $V_{ub}$) is especially 
important to the Standard Model description
of CP-violation. 
If it were zero, there would be no CP-violation from
the KM matrix elements 
({\it i.e.} within the Standard Model), 
and we have to seek
for other source of CP violation in 
$K_{L} \rightarrow \pi\pi$.
Here  we study  the ratio $V_{td} \over V_{ts}$. 
In the Standard Model with the
unitarity of KM matrix, 
$V_{ts}$ is approximated by $-V_{cb}$,
which is directly measured by semileptonic $B$ decays.
There are already several ways to determine $V_{td}$ available
in the literature:
\begin{itemize}
\item
$|V_{td}|$ can be indirectly extracted through
$B_d - \overline{B}_d$ mixing.
However, in $ B_d -\overline{B}_d$ mixing 
the large uncertainty of the hadronic
matrix elements prevents us to extract the element of KM with
good accuracy.  
\item
A better extraction of $|{V_{td} \over V_{ts}}|$ can be made
if $ B_s -\overline{B}_s$ is measured as well, because
the ratio $f_{_{B_d}}^2 B_{_{B_d}}/f_{_{B_s}}^2 B_{_{B_s}}$
can be much better determined.
\item
The determination of $|{V_{td} \over V_{ts}}|$ from the ratios
of rates of several hadronic two-body $B$ decays, such as  
$\Gamma(B^0 \to \overline{K}^{*0} K^0)
   /\Gamma(B^0 \to \phi K^0)$, 
$\Gamma(B^0 \to \overline{K}^{*0} K^{*0})
   /\Gamma(B^0 \to \phi K^{*0})$, 
$\Gamma(B^+ \to \overline{K}^{*0} K^+) 
   /\Gamma(B^+ \to \phi K^+)$, and 
$\Gamma(B^+ \to \overline{K}^{*0} K^{*+})
   /\Gamma(B^+ \to \phi K^{*+})$ 
has been also proposed in Ref. \cite{gronau}.
\item
$V_{td}$ can be determined from $K^+\to\pi^+\nu\bar\nu$ 
decay within $\sim 15 \%$ theoretical uncertainties
with the branching fraction ${\cal B} \sim 10^{-10}$ \cite{buras2}.
\end{itemize}

Why do we discuss yet another method to determine $V_{td}$? 
The main reason
we are interested in $B$ physics is that this area is very likely to
yield information about new physics beyond the Standard Model. 
We expect that new physics will influence 
experimentally measureable quantities in different ways. For example,
most of us expect that $\Delta B=2$ transition is more sensitive
to new physics than the decay rates. New physics may couple
differently to $K$ mesons compared to $B$ mesons. 
Therefore, it is essential
to determine the KM matrix elements in as many different methods
as possible.

In this paper,
we propose another method  to determine 
${|V_{td}| \over  |V_{ts}|}$ 
precisely from the decay distributions
$\mbox{\dBs}$ and  $\mbox{\dBd}$, where $s$
is invariant mass-squared of final lepton pair, $l^+ l^-$. 
In the decays of 
$\mbox{\bxqll} (q=d,s)$, the short distance (SD)
contribution  comes from the top quark loop diagrams 
and the long distance (LD) contribution  comes 
from the decay chains due to intermediate charmonium states.
Therefore, the former (SD) amplitude is proportional to 
${V_{tq}}^* V_{tb}$, and the latter (LD) proportional to
${V_{cq}}^* V_{cb}$. If the invariant mass-squared of 
$l^+ l^-$ is away from the peaks of the  
charmoniuum resonances ($J /\psi$ and $\psi'$), 
the SD contribution is dominant, while on the peaks
of the resonances the LD contribution is dominant,
and therefore we expect that in the $SU(3)$ symmetry limit 
the ratio
\def\dBs{$ {d{\cal{B}} \over ds} [\mbox{\bxsll}] $}
\begin{displaymath}
{dR \over ds} \equiv {d{\cal{B}} \over ds}
  (B \rightarrow X_d l^+ l^-)/{d{\cal{B}} \over ds}
  (B \rightarrow X_s l^+ l^-)
\end{displaymath}
becomes 
\begin{eqnarray}
{1 \over \lambda^2}{dR(s) \over ds} 
   & \rightarrow&  (1-\rho)^2+\eta^2
   \quad (s \rightarrow 1 \rm{GeV}^2) ~, \nonumber \\
   &\rightarrow & \qquad 1  \qquad  \qquad  ( s
\rightarrow  {m_{_{J/\psi}}}^2) ~,
\end{eqnarray}
where $\lambda$ is 
$sin \theta_c$, $\theta_c$= Cabibbo angle, 
and for $\rho$, $\eta$, see Eq. (\ref{eq8}).
In the intermediate region, there is a
characteristic interference beteen the LD contribution and
the SD contribution, which requires the detailed study of 
the distributions.
By focusing on the region, 
$1 ({\rm GeV}^2)  \le s  < m_{J/\psi}^2 $,        
we can study 
$|{V_{td} \over V_{ts}}|$ from the experimental ratio 
of the distributions ${dR \over ds}$.

We stress    
the advantages to use  the inclusive semileptonic  decays:
\begin{itemize}
\item{If dilepton invariant mass-squared $s$ is away
from the peaks of the  possible intermediate states,  
$J/\psi$, $\psi'$, $\rho$, $\omega$, and {\it etc.}, 
the short distance contribution due to top
quark loop is dominant. 
Therefore, we may extract very precisely the
combination of $|{V_{td} \over  V_{ts}}|$ from the decay
distributions within the range 
$ 1 ({\rm GeV}^2) \le s < {m_{J/ \Psi}}^2$ 
without any theoretical uncertainties from
the LD hadronic matrices, 
unknown KM matrix elements, and {\it etc}. } 
\item{Inclusive decay distributions are theoretically 
well predicted by
heavy quark mass expansion away from the phase space
boundary.  The leading order of the expansion agrees with 
the parton model result. Furthermore,  non-perturbative 
$1/{m_b}^2$ power corrections of QCD 
can be easily incorporated.}
\end{itemize}

This paper is organized as follows.
In Section 2,  we present the analytic formulae for
$\mbox{\dBs}$ and $\mbox{\dBd}$. 
We also show in detail the decay
distribution $\mbox{\dBs}$ including the uncertainties coming
from top quark mass $m_t$ and the scale of  renormalization
group $\mu$. 
In Section 3, after brief discussion 
on the limit of KM elements coming  from 
$B_d- \overline{B}_d $ mixing and $\mbox{\BXu}$,
we  study the ratio of the decay
distributions, and show how we may extract 
$ |{V_{td} \over V_{ts}}|$.

\section{\bf Effective Hamiltonian for \bqll
and their differential decay rates.}
 
In this Section,  the differential decay rates 
for $ \mbox{\bxqll} (q=d,s)$ 
are shown including both LD and SD effects as well as
${\Lambda_{QCD}}^2 / {m_b}^2$ power corrections.
(See \cite{ks} for $b \rightarrow d l^+ l^-$,
   \cite{long} for $b \rightarrow s l^+ l^-$,
and \cite{fl}, \cite{ahhm} for $\mbox{\bxsll}$ including the 
$1 / {m_b}^2$ power corrections.)
We also show in detail how the differential 
decay rate varies by changing the input parameters $m_t$, 
and  the renormalization scale $\mu$.
In Table 1, we summarize all the values of the input parameters 
used in our numerical calculations of decay rates.
We use the central values for those input parameters, 
unless otherwise specified.

The effective Hamiltonian for \bqll $(q=d,s)$ is given as
\begin{eqnarray}
{\cal H}_{eff}
  &=& -\frac{4 G_F}{\sqrt{2}} V_{tq}^*
V_{tb}
\left[ \sum_{i=1}^{10} C_i   O_i  \right] \nonumber \\
&+&\frac{4 G_F}{\sqrt{2}} V_{uq}^*
V_{ub}
\left[ C_1  ({O_1}^{u}-O_1) + C_2 ({O_2}^{u}-O_2) \right],
\label{heffbsll}
\end{eqnarray}
where $V_{ij}$ are the KM matrix elements.
The operators are given as
\begin{eqnarray}
 O_1 &=& (\bar{q}_{L \alpha} \gamma_\mu b_{L \alpha})
               (\bar{c}_{L \beta} \gamma^\mu c_{L \beta}),
 \nonumber \\
 O_2 &=& (\bar{q}_{L \alpha} \gamma_\mu b_{L \beta})
               (\bar{c}_{L \beta} \gamma^\mu c_{L \alpha}),
 \nonumber \\
 O_3 &=& (\bar{q}_{L \alpha} \gamma_\mu b_{L \alpha})
               \sum_{q'=u,d,s,c,b}
               (\bar{q'}_{L \beta} \gamma^\mu q'_{L \beta}),
 \nonumber \\
 O_4 &=& (\bar{q}_{L \alpha} \gamma_\mu b_{L \beta})
                \sum_{q'=u,d,s,c,b}
               (\bar{q'}_{L \beta} \gamma^\mu q'_{L \alpha}),    
 \nonumber \\
O_5 &=& (\bar{q}_{L \alpha} \gamma_\mu b_{L \alpha})
               \sum_{q'=u,d,s,c,b}
               (\bar{q'}_{R \beta} \gamma^\mu q'_{R \beta}),    
 \nonumber \\
O_6 &=& (\bar{q}_{L \alpha} \gamma_\mu b_{L \beta})
                \sum_{q'=u,d,s,c,b}
               (\bar{q'}_{R \beta} \gamma^\mu q'_{R \alpha}),      
 \nonumber \\
O_7 &=& \frac{e}{16 \pi^2}
 \bar{q}_{\alpha} \sigma_{\mu \nu} (m_b R + m_q L) b_{\alpha}
                F^{\mu \nu},
                                     \nonumber \\
 O_8 &=& \frac{g}{16 \pi^2}
    \bar{q}_{\alpha} T_{\alpha \beta}^a 
                \sigma_{\mu \nu} (m_b R + m_q L)  
          b_{\beta} G^{a \mu \nu},  \nonumber \\
 {O_1}^{u} &=&  (\bar{q}_{L \alpha} \gamma_\mu b_{L \alpha})
               (\bar{u}_{L \beta} \gamma^\mu u_{L \beta}),
\nonumber \\
 {O_2}^{u} &=&  (\bar{q}_{L \alpha} \gamma_\mu b_{L \beta})
               (\bar{u}_{L \beta} \gamma^\mu u_{L \alpha}),
\nonumber \\    
 O_9 &=& \frac{e^2}{16 \pi^2} \bar{q}_\alpha 
             \gamma^{\mu} L b_\alpha
\bar{\ell} \gamma_{\mu} \ell , \nonumber\\
 O_{10} &=& \frac{e^2}{16 \pi^2} \bar{q}_\alpha \gamma^{\mu} L
b_\alpha \bar{\ell} \gamma_{\mu}\gamma_5 \ell ,
\end{eqnarray}
where $L$ and $R$ denote chiral projections, 
$L(R)=1/2(1\mp \gamma_5)$.
We use the Wilson coefficients given in the
literature (see, for example, \cite{buras}).

With  the effective Hamiltonian in Eq.~(\ref{heffbsll}),
the matrix element for the decays 
$ \mbox{\bqll} (q=d,s)$ can be written as 
\begin{eqnarray} 
	{\cal M (\mbox{\bqll})} & = & 
     \frac{G_F \alpha}{\sqrt{2} \pi} \, V_{tq}^\ast V_{tb} \, 
	\left[ \left( C_{9q}^{\mbox{eff}} - C_{10} \right) 
		\left( \bar{q} \, \gamma_\mu \, L \, b \right)
		\left( \bar{l} \, \gamma^\mu \, L \, l \right) 
                \right. \nonumber \\
        & & \left.
                \; \; \; \; \; \; \; \; \;
                \; \; \; \; \; \; \; \; \;
		+ \left( C_{9q}^{\mbox{eff}} + C_{10} \right) 
		\left( \bar{q} \, \gamma_\mu \, L \, b \right)
		\left( \bar{l} \, \gamma^\mu \, R \, l \right)  
		\right. \nonumber \\
	& & \left. 
                \; \; \; \; \; \; \; \; \;
                \; \; \; \; \; \; \; \; \;
 - 2 C_7^{\mbox{eff}}\left(\bar{q} \, i \, \sigma_{\mu \nu} \, 
	\frac{q^\nu}{q^2} (m_q L + m_b R) \, b \right)  
		\left( \bar{l} \, \gamma^\mu \, l \right) 
		\right] \, , 
	\label{eqn:hamiltonian}
\end{eqnarray} 
where $C_{9q}^{\mbox{eff}}$ is given by
\begin{equation}
C_{9q}^{\mbox{eff}} (\hat{s}) \equiv C_9
\left\{ 1 + \frac{\alpha_s(\mu)}{\pi}
                \omega(\s) \right\} +
{Y_{SD}}^q (\hat{s})
 + {Y_{LD}}^q (\hat{s}) ~,
\label{eqn:Y}
\end{equation}
where $q^\nu$ is the four momentum of $ l^+ l^-$, 
$s=q^2$, and  $\hat{s}=s / {m_b}^2$.
The function ${Y_{SD}}^q (\hat{s})$ is 
the one-loop matrix element of $ O_9$,
and  ${Y_{LD}}^q (\hat{s})$ is the LD contributions 
due to the vector mesons mesons $J/\psi$, $\psi^\prime$ 
and higher resonances. 
The function $\omega(\hat{s})$ represents the
$O(\alpha_s)$ correction from the one-gluon exchange 
in the matrix element of $O_9$, 
and is given in our Appendix. The two functions
${Y_{SD}}^q$ and ${Y_{LD}}^q$
are written as
 \begin{eqnarray}
        {Y_{SD}}^q (\s) & = & g(\mc,\s)
                \left(3 \, C_1 + C_2 + 3 \, C_3
                + C_4 + 3 \, C_5 + C_6 \right)
\nonumber \\
        & & - \frac{1}{2} g(1,\s)
                \left( 4 \, C_3 + 4 \, C_4 + 3 \,
                C_5 + C_6 \right) \nonumber \\
        & & - \frac{1}{2} g(0,\s) \left( C_3 +   
                3 \, C_4 \right) \nonumber \\
        & &     + \frac{2}{9} \left( 3 \, C_3 + C_4 +
                3 \, C_5 + C_6 \right) \nonumber \\
        & &   -{\Vuq^* \Vub \over \Vtq^* \Vtb} (3 C_1 +C_2)
               (g(0 , \s)-g(\mc, \s)) ~,\label{eqn:y}
\end{eqnarray}
\begin{eqnarray}
         {Y_{LD}}^q(\s) & = & 
                      \frac{3}{\alpha^2} \kappa \,
       \left( -{ \Vcq^* \Vcb \over \Vtq^* \Vtb}  C^{(0)}
              -{\Vuq^* \Vub \over \Vtq^* \Vtb} ( 3 \, C_3
              + C_4 + 3 \, C_5 + C_6 )\right) \nonumber \\
	& & \times \sum_{V_i = \psi(1s),..., \psi(6s)}
      \frac{\pi \, \Gamma(V_i \rightarrow l^+ l^-)\, M_{V_i}}{
      {M_{V_i}}^2 - \s \, {m_b}^2 - i M_{V_i} \Gamma_{V_i}} ~,
                \label{LDeq} 
\end{eqnarray}
where
the function $g(\hat{m_c}={m_c \over m_b},\hat{s})$, 
$g(1, \hat{s})$, and
$g(0, \hat{s})$ represent  
$c$ quark, $b$  quark and $u,d,s$ quark loop  contributions,
respectively. The two functions $g(z, \hat{s})$ and 
$g(0, \hat{s})$ are  given \cite{buras} in our Appendix.
In Eq. (\ref{LDeq}),
$C^{(0)} \equiv 3 C_1  + C_2  + 3 C_3 + 
	C_4  + 3 \, C_5  + C_6  $, 
and the first two terms are dominant, as can be seen from 
the Table $2$. 

It is convenient to write the relevant combinations of KM 
in terms of the Wolfenstein parametrization.
In the following, in addition to  $A$ and $ \lambda$,
we choose $ (1-\rho)^2 + \eta^2$ and
$\phi_1 =arg  (-{ {V_{cd}V_{cb}^*} \over {V_{td}V_{tb}^*}})$  
as independent variables, then we have
\begin{eqnarray}
 V_{ts}^* V_{tb}&\simeq &\lambda^2 A ~, \nonumber \\
\frac { V_{cs}^* V_{cb} }{ V_{ts}^* V_{tb} } &\simeq&  -1 ~,
\nonumber  \\ 
\frac { V_{us}^* V_{ub}}{ V_{ts}^* V_{tb} } 
                &\simeq&  O( \lambda^2) ~,
\nonumber \\
 V_{td}^* V_{tb} &\simeq & \lambda^3 A (1- \rho +i \eta) ~,
\nonumber \\
\frac { V_{cd}^* V_{cb}}{ V_{td}^* V_{tb} }
& \simeq & -\frac {1} {1- \rho +i \eta} ~,
\nonumber \\
 & = &  - \frac {\exp (-i \phi_1)} {\sqrt{(1-\rho)^2 +
\eta^2} } ~, \nonumber \\
\frac { V_{ud}^*  V_{ub}} {  V_{td}^* V_{tb} }
& \simeq & \frac { \rho - i \eta} {1- \rho + i \eta} ~,
\nonumber \\
&=& -1+ \frac {\exp (-i \phi_1)} {\sqrt{(1-\rho)^2 +
\eta^2}} ~.
\label{eq8}
\end{eqnarray}  
In the SD contribution of  $\mbox{\bxsll}$, 
the $u$-quark loop contribution is neglected
due to the smallness of the combination $ V_{us}^* V_{ub}$ 
compared with $ V_{cs}^*  V_{cb}  \simeq - V_{ts}^* V_{tb} $,
while  in  $ \mbox{\bxdll} $ the term
which is proportional to $ V_{ud}^* V_{ub} $ is maintained. 
In the LD contribution, there is a contribution
coming from the gluonic penguin amplitude which
is proportional to 
$ \frac {V_{ud}^* V_{ub}} {V_{td}^* V_{tb} } 
                        (3 C_3 + C_4 + 3 C_5 + C_6)$. 
This contribution is neglected because of the smallness
of the Wilson coefficient $ (3 C_3 + C_4 + 3 C_5 + C_6) $
compared with $C^{(0)}$.
We adopt $\kappa = 2.3 $ \cite{LW96} to reproduce the
rate of the decay chain  
$ B \rightarrow X_s J /\psi \rightarrow X_s l^+ l^-$.
Note that the data determines only the combination, 
$\kappa \, C^{(0)} = 0.88$.

By combining both SD and LD contributions as well as
non-perturbative $1 / {m_b}^2$ power corrections, 
the differential decay rate for $\mbox{\bxqll} (q=d,s)$ becomes
\begin{eqnarray}
& &\frac{{\rm d}{\cal B}}{{\rm d}\s}=
   2         \frac {{\vert V_{tq}^*
                    V_{tb}\vert}^2}{\absvcb^2}
      \; {\cal B}_0
                \left\{
                  \left[
                \frac{2}{3} \u(\s,\mq)
((1 - \mq^2)^2 + \s (1+\mq^2) -2 \s^2)
\right. \right.
\nonumber \\
        & &
+   \frac{1}{3} (1 -4 \mq^2 + 6 \mq^4 -4 \mq^6 + \mq^8
-\s
+ \mq^2 \s +
\mq^4 \s - \mq^6 \s
\nonumber \\
& &
 -3 \s^2 -2 \mq^2 \s^2 -3 \mq^4 \s^2
+ 5 \s^3 +5 \mq^2 \s^3-2 \s^4 ) \frac{\lo}{ \u(\s,\mq)}
\nonumber \\
  & &   + \left( 1 -8 \mq^2 + 18 \mq^4 -16 \mq^6 + 5 \mq^8 -\s
-3 \mq^2 \s + 9 \mq^4 \s -5 \mq^6 \s -15 \s^2
\right. \nonumber \\
& &
\left. \left.-18 \mq^2 \s^2
-15 \mq^4 \s^2 + 25 \s^3
 + 25 \mq^2 \s^3 -10 \s^4 \right) \frac{\lt}{ \u(\s,\mq)}
                        \right]
         \left( |C_9^{\mbox{eff}}|^2 + |C_{10}|^2 \right)
                \nonumber \\
        &  &
            +   \left[
      \frac{8}{3} \u(\s, \mq) \left(2 (1+\mq^2)(1-\mq^2)^2-
            (1+14 \mq^2 +\mq^4) \s -(1+\mq^2) \s^2 \right)
                \right.
                \nonumber \\
        & & 
   + \frac{4}{3} (2 - 6 \mq^2 + 4 \mq^4 +4 \mq^6 -6 \mq^8+
 2 \mq^{10}     
\nonumber \\ & &
-5 \s -12 \mq^2 \s + 34 \mq^4 \s -12 
                     \mq^6 \s -5 \mq^8 \s + 3 \s^2
\nonumber \\
& &
 + 29 \mq^2 \s^2 + 29 \mq^4 \s^2
+3 \mq^6 \s^2+ \s^3 -10 \mq^2 \s^3 +
                   \mq^4 \s^3-\s^4-\mq^2 \s^4)
  \frac{\lo}{ \u(\s,\mq)}
\nonumber \\
& &+ 4  \left(-6 + 2 \mq^2 
 + 20 \mq^4 -12 \mq^6 - 14 \mq^8 +10 \mq^{10}
 +
  3 \s+   16 \mq^2 \s +
 62 \mq^4 \s  \right. \nonumber \\ & & -56 \mq^6 \s
 -25 \mq^8 \s + 3 \s^2
 + 73 \mq^2 \s^2 + 101 \mq^4 \s^2
+15 \mq^6 \s^2+ 5 \s^3 
-26 \mq^2 \s^3+
\nonumber \\ & & \left.\left.
5 \mq^4 \s^3 -5 \s^4-5 \mq^2 \s^4 \right) 
            \frac{\lt}{ \u(\s,\mq)}
         \right] \frac{|C_7^{\mbox{eff}}|^2}{\s}
                 +      \left[
8 \u(\s, \mq) \left((1-\mq^2)^2 \right.\right. \nonumber\\ & &
\left. -(1+\mq^2) \s \right)
            + 4 ( 1 - 2 \mq^2 +\mq^4  - \s-\mq^2 \s)
\; \u(\s,\mq) \;  \lo
\nonumber \\
 & & + 4 \left( -5 +30\mq^4-40 \mq^6 +15 \mq^8 -\s + 21
\mq^2 \s +25 \mq^4 \s -45 \mq^6 \s + 13 \s^2
\right.  \nonumber \\
& &
\left. \left. \left.
 + 22 \mq^2 \s^2
+45 \mq^4 \s^2 -7 \s^3-15 \mq^2 
                    \s^3 \right)\frac{\lt}{ \u(\s,\mq)}
  \right] Re(C_9^{\mbox{eff}}) \, C_7^{\mbox{eff}}
      \right\}
\, .
\label{eq9}
\end{eqnarray}

\begin{figure}[htb]
\begin{center}
\leavevmode\psfig{file=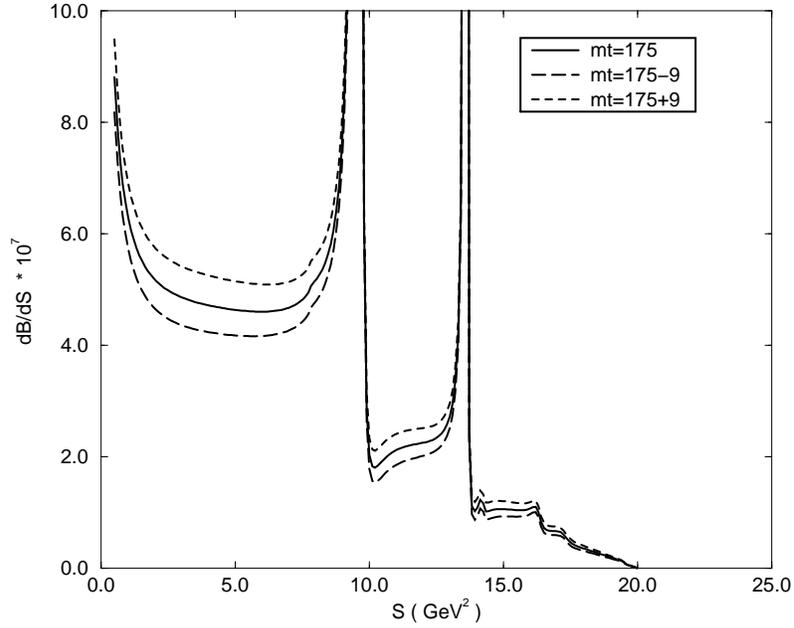,width=12cm,angle=-90}
\end{center}
\caption{
Dilepton invariant mass spectrum, 
${{\rm d}{\cal B} \over {\rm d} s}(B \to X_s e^+ e^-) 
                 \times 10^7$.  
The unit of $s$ is $ {\rm (GeV^2 ) }$. 
The solid line corresponds to $m_t =175~(\rm{GeV})$.  
The short-dashed line corresponds to $m_t =175+9$. 
The long-dashed line corresponds to $m_t =175-9.$ }
\label{fig:1}
\end{figure}

\begin{figure}[htb]
\begin{center}
\leavevmode\psfig{file=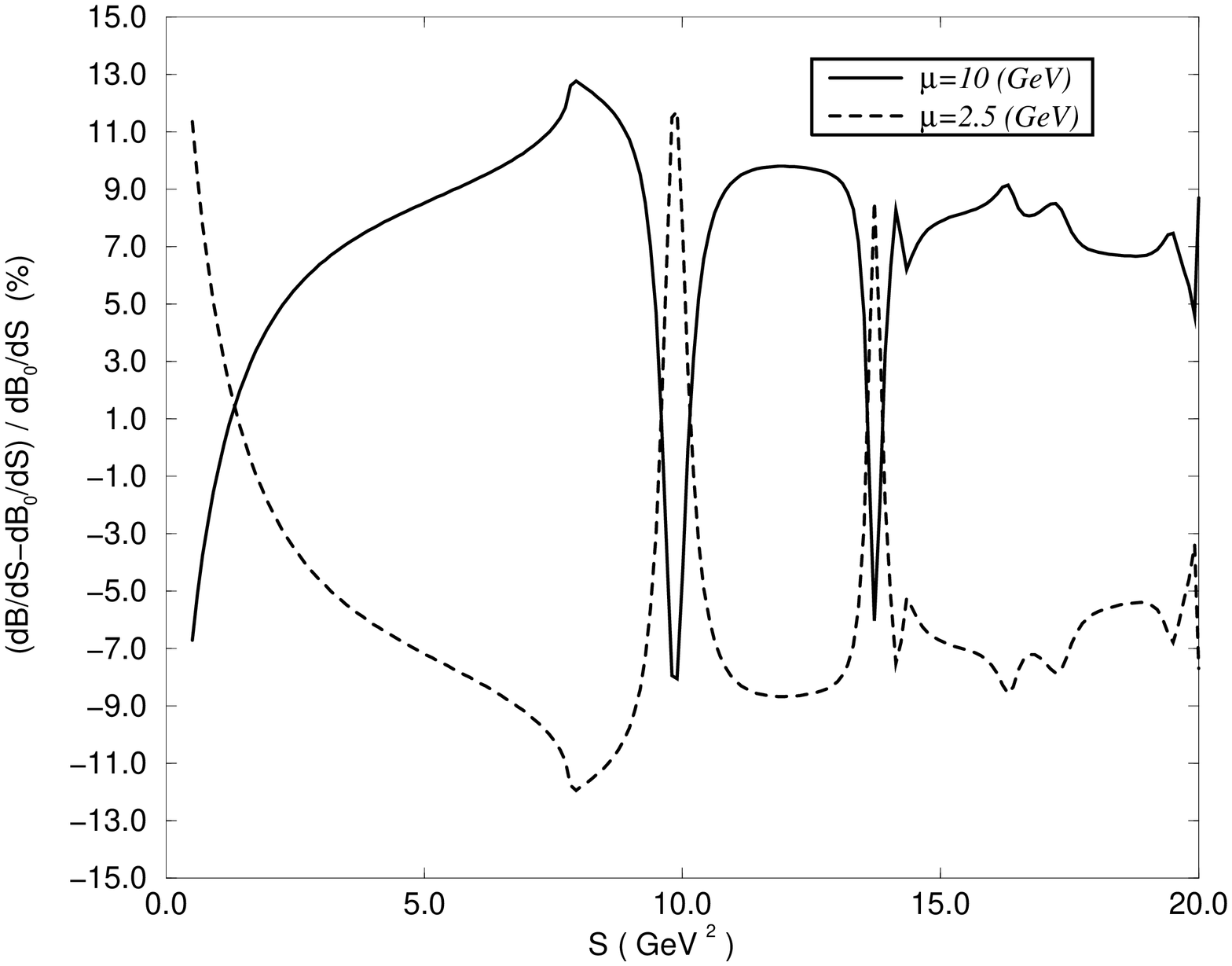,width=12cm,angle=-90}
\end{center}
\caption{The $\%$ variation of dilepton invariant mass
spectrum 
defined as
$({{\rm d}{\cal B}(\mu) \over {\rm d} s}
-{{\rm d}{\cal B}_0 \over {\rm d} s})/ {{\rm d}{\cal B}_0
\over {\rm d} s}$;  ${\cal B}_0={\cal B}(\mu=5 \mbox{(GeV)}$).
The solid  line corresponds to $\mu =10$ and   
the dashed line corresponds to $\mu =2.5.$ }
\label{fig:2}
\end{figure}

We explain some aspect of the  expression, (\ref{eq9}):
\begin{itemize}
\item{
The branching ratio  is normalized by 
${\cal B}_{sl}$ of  decays $B \to (X_c,X_u) \ell
\nu_\ell$. We separate a combination of the KM factor 
 $\frac {{\vert V_{tq}^* V_{tb}\vert}^2}{\absvcb^2}$      
due to top-quark loop from the normalization factor  
${\cal B}_0$. 
The normalization constant ${\cal B}_0$ is
\begin{equation}
    {\cal B}_0  \equiv
       {\cal B}_{sl} \frac{3 \, \alpha^2}{16 \pi^2} 
           \frac{1}{f(\mc) \kappa(\mc)}
                \, ,
\label{eqn:seminorm}
\end{equation}
where $f(\mc)$ is a phase space factor, 
and $\kappa(\mc)$ accounts for both the $O(\alpha_s)$ QCD 
correction to the semi-leptonic decay  
width and the leading order
$(1/m_b)^2$ power correction,
They are given in our Appendix explicitly.}
\item{The Wilson coefficients depend on the top quark mass
$m_t$, the renormalization scale $\mu$, and $\Lambda_{QCD}$.
Their dependences are studied in Tables 2, 3 and 4.
As the value of $ \mu $ ($ \Lambda_{QCD}$) decreases 
(increases) , $|C_{7}|$ and $ C_{9}$ are getting larger, while 
$C_{10}$ is independent on $\mu$. 
As $m_t$ increases,  
$|C_{7}|$, $C_{9}$ and $|C_{10}|$ all become larger.
The matrix elements also  depend on the renormalization  
scale $\mu$, and their dependence 
is partially cancelled by the given dependence of the Wilson
coefficients. The overall dependence can be studied with 
the differential rate. 
In Figs. \ref{fig:1} and \ref{fig:2}, we  show  
the dependence of the differential rate on $m_t$ and $\mu$.
As the value of $m_t$ increases  from 166 (GeV) to  184 (GeV),
the differential rate increases about $20 \%$
at  $s \sim 5$ (GeV$^2$), as shown in Fig. \ref{fig:1}.
At around the same region of $s$,
by changing $\mu$ from 2.5 (GeV) to 10 (GeV) 
the differential rate increases  about  $20 \%$,
as shown in Fig. \ref{fig:2}.
This observation is  consistent with the result of 
Ref. \cite{buras}, in which only the SD contribution 
has been analysed.
}
\item{The differential decay rate (\ref{eq9}) 
is not a simple parton model result. 
It  contains non-perturbative $1 / {m_b}^2$ power corrections, 
which are denoted in Eq. (\ref{eq9}) by the terms proportional to
$ \hat{\lambda}_1$ and $ \hat{\lambda}_2$. 
The  parameters $ \hat{\lambda}_1$ and $ \hat{\lambda}_2$
are related to the  matrix elements of the higher derivative
operators of heavy quark effective theory
\cite{fl}, \cite{georgi};
\begin{eqnarray}
 \left< B \left| \bar{h} \, (i \, D)^2 \, h \right| B\right>  
 & \equiv & 2 \, M_B \, \loo = 2  M_B {m_b}^2 \hat{\lambda}_1
           \, , \nonumber \\
  \left< B \left| \bar{h} \, \frac{-i}{2} \sigma^{\mu \nu}
     \, G_{\mu \nu} \, h \right| B \right>
 & \equiv & 6 \, M_B \, \lto = 6 M_B {m_b}^2 \hat{\lambda}_2
                \, ,
\label{hqetpar}
\end{eqnarray}
where $B$ denotes the pseudoscalar $B$ meson, 
$D_\mu$ is the covariant derivative, 
and $G_{\mu \nu}$ is the QCD field strength tensor.
}
\end{itemize}

\section{\bf The extraction of 
           $\left|{V_{td} \over V_{ts}}\right|$ }

Presently  we can  constrain the value of 
$ (1-\rho)^2 + \eta^2$ from $B_d - \overline{B}_d$ mixing,  
while from   $ B \rightarrow X_u l \bar{\nu}$ 
we obtain some limit  on  $\rho^2 + \eta^2$.
Using those given limits,
we can show how  the ratio $ {1 \over \lambda^2} {dR \over ds}$ 
depends on input KM values,
$ (1-\rho)^2 + \eta^2$ and $\phi_1~({\rm or}~\beta)$.

\begin{figure}[htb]
\begin{center}
\leavevmode\psfig{file=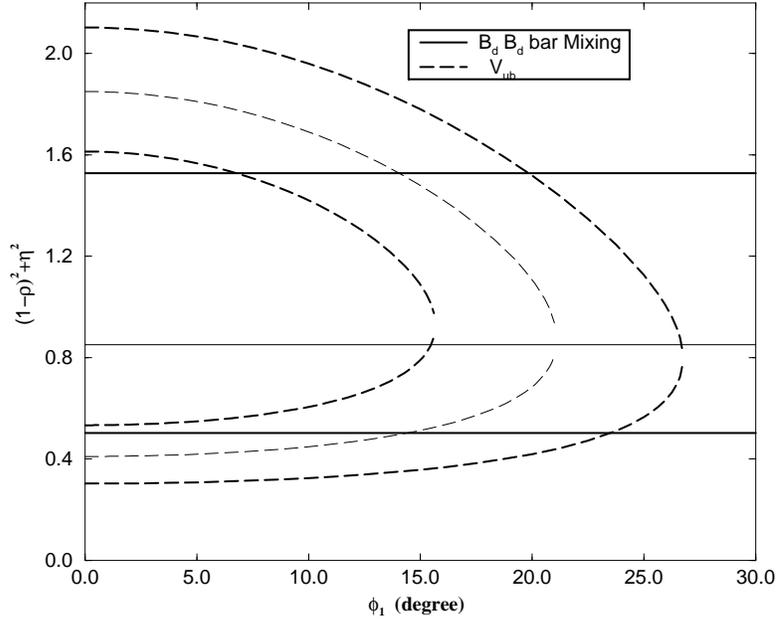,width=12cm,angle=-90}
\end{center}
\caption{ The limit on $(1-\rho)^2+\eta^2$ and   
$ \phi_1~({\rm or}~\beta)$.
The horizontal axis corresponds to 
$\phi_1$, and the unit is degree. 
The vertical axis corresponds to $(1-\rho)^2+\eta^2$.
The thin dashed line and the thin solid line 
are  obtained from the central values
of $|V_{ub}|$ and  $|V_{td}|$ respectively.
The thick solid lines are obtained from the allowed range of
$|V_{td}|$ from $B_d \bar{B_d}$ mixing. The thick dashed lines
are obtained from the allowed range of $|V_{ub}|$. }
\label{fig:3}
\end{figure}

As is well known,  $ |V_{tb}^* V_{td}| $ are  written with 
$\Delta M_d$ of  $B_d - \overline{B}_d$ mixing. It reads as 
\begin{eqnarray}
|V_{tb}^* V_{td}|&=& A \lambda^3 \sqrt{(1-\rho)^2 + \eta^2}
 \nonumber \\ &=& 
 \left[ \frac { 6 \pi^2} {G_F^2 \eta_{_{QCD}}
M_W^2 M_B} \right]^{1/2} \left[ \frac{ {\Delta {M_d}}^{1/2}}
{ B_{_{B_d}}^{1/2} f_{_{B_d}} |E(x_t)|^{1/2}} \right] , 
 \end{eqnarray}
where $E(x_t)$ is 
the Inami-Lim function \cite{cslim} of the box diagram,
and  $x_t \equiv \frac{ {m_t}^2} {{m_W}^2}$.
It has the value 
$|E(x_t)|=2.58 $ for $m_t= 175$ (GeV) and it varies  
from $2.38$ to $2.78$ as $m_t$ varies from 
$166$ (GeV) to $184$ (GeV).
We use the following values;
$\eta_{_{QCD}}=0.55$, $m_t=175$ (GeV), 
and the experimental constraints 
\begin{eqnarray} 
\Delta M_d &=& 0.474 \pm 0.031 (ps^{-1}) ~, \nonumber\\
\left|{V_{ub} \over V_{cb}}\right| &=& 
                    \lambda \sqrt{\rho^2 + \eta^2} 
=0.08 \pm 0.02 ~.
\label{eq14}
\end{eqnarray}
Then, those constraints (\ref{eq14}) 
are translated into the limits of 
$(1-\rho)^2 + \eta^2$
and $\sqrt{\rho^2 + \eta^2}$ as
\begin{eqnarray}
(1-\rho)^2 + \eta^2&=&
 0.85(1 \pm 0.15) 
     \left\{ 0.2 / (B_{_{B_d}}^{1/2} f_{_{B_d}}({\rm GeV})) 
 \right \}^2 \nonumber \\
 &=&  [0.59 \pm 0.09 , 1.3 \pm 0.20 ] ~, \nonumber \\ 
\sqrt{\rho^2 + \eta^2} &=& 0.36 \pm 0.09 ~.
\end{eqnarray}
Here we used 
$ B_{_{B_d}}^{1/2} f_{_{B_d}} = 0.2 \pm 0.04 (\rm{GeV}) $ 
to get the range of 
$(1-\rho)^2 + \eta^2 \approx [0.5, 1.5]$.  
We also used $ A \lambda^2 =0.041 (\pm 0.003)$,
and $\lambda=0.2205$.
In Fig. \ref{fig:3}, 
we show the present limit in  the plane of
$(\phi_1,~ (1-\rho)^2+\eta^2)$.
The horizontal axis corresponds to 
$\phi_1~({\rm or}~\beta)$, and the unit is
degree. The vertical axis corresponds to
$(1-\rho)^2+\eta^2$.
The thin dashed line is obtained from the central values of 
$|V_{ub}|$ and the thin solid line is obtained from $|V_{td}|$. 
The central value corresponds to  
$(\phi_1,~ (1-\rho)^2+\eta^2) \approx (20^o,~0.85)$.
 
\begin{figure}[htb]
\begin{center}
\leavevmode\psfig{file=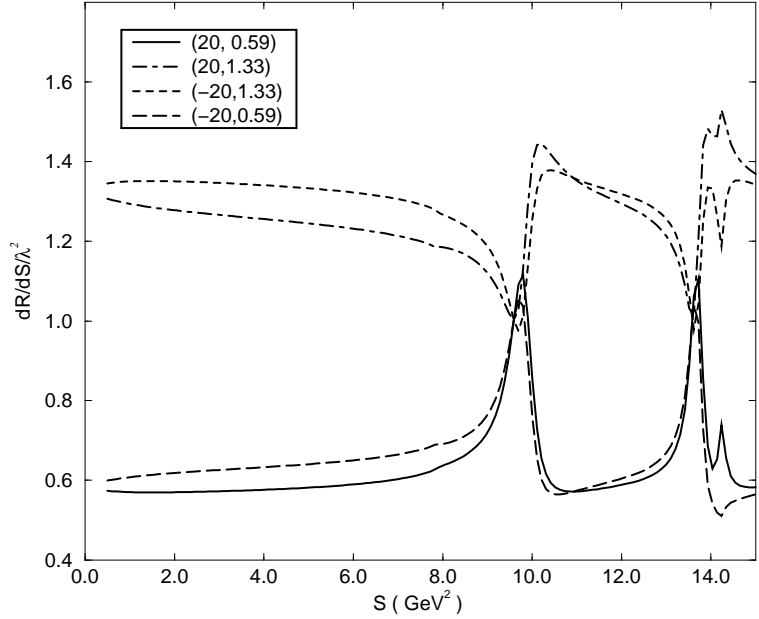,width=12cm,angle=-90}
\end{center}
\caption{ The ratio ${1 \over \lambda^2} 
{dR \over ds}$ versus $s$~(GeV$^2$)
with two different input values of $(1-\rho)^2+\eta^2$.
The solid curve correspond to 
$(\phi_1,~(1-\rho)^2+\eta^2) =  (20^o,~0.59)$, 
and the dot-dashed curve  corresponds to  $(20^o,~1.33)$.  
The ratios for CP conjugate process 
${\overline B} \rightarrow {\bar X_q} l^+ l^- $
are  denoted by the dashed curve  
for $(1-\rho)^2+\eta^2=1.33$,
and by the long-dashed curve for $(1-\rho)^2+\eta^2=0.59$.
They are obtained by reversing the sign of $\phi_1$ in the
corresponding  $ B \rightarrow  X_q l^+ l^- $ process; 
$i.e.$, $\phi_1 \rightarrow -\phi_1$.}
\label{fig:4}
\end{figure}

Now let us consider the ratio 
${1 \over \lambda^2} {dR(s)\over ds}$.
In the SU(3) limit, 
we expect  that this ratio  approaches to 
the input value of  $(1-\rho)^2+\eta^2$ 
for the dileptonic invariant mass-squared $s$ 
far below  the peak of charmonium resonances.
If $s$ is on the peak of resonances, the ratio becomes 1. 
In Fig. \ref{fig:4}, 
we show the ratio  ${1 \over \lambda^2} {dR(s)\over ds}$ 
for two  sets of the assumed input values of 
$(\phi_1,~ (1-\rho)^2+\eta^2)$,
one of whose $(1-\rho)^2+\eta^2$ is 0.59, and the other
is 1.33. They correspond to their small (or large) 
values allowed from  the present experimental result of 
$B_d - \overline{B}_d$ mixing. 
The solid curve corresponds to
$(\phi_1,~(1-\rho)^2+\eta^2) =  (20^o,~0.59)$, 
and the dot-dashed curve
corresponds to  $(20^o,~1.33)$.
The ratios for the CP conjugate process 
${\overline B} \rightarrow {\bar X_q} l^+ l^- $ 
are  also shown,
denoted by the long-dashed curve  for 
$(1-\rho)^2+\eta^2=1.33$,
and by the dashed curve for $(1-\rho)^2+\eta^2=0.59$.
They are obtained by  reversing the sign of $\phi_1$ in
the corresponding $B \rightarrow  X_q l^+ l^-$ process; 
{\it i.e.}, $\phi_1 \rightarrow -\phi_1 $. 
They are labeled as $(-20^o,~1.33)$ and  $(-20^o,~0.59)$
in Figure \ref{fig:4}. 
 
To summarize the numerical results of 
Fig. \ref{fig:4} and Section 3:
\begin{itemize}
\item{The predicted ratio at low invariant mass region 
$s \simeq 1$ (GeV$^2$) is very near to the our assumed 
input value of $(1-\rho)^2 + \eta^2$,
while on the peak of the resonances 
$J/\psi$, $\psi'$, this ratio
becomes almost 1, as we expected, {\it i.e.}
\begin{eqnarray}
{1 \over \lambda^2}{dR(s) \over ds} &\rightarrow&  
     \quad\quad 1 \quad\quad\quad\quad 
(s \sim {m_{J/\psi}}^2,{m_{\psi'}}^2,..) ~, \nonumber \\
&\rightarrow &  
 (1-\rho)^2+\eta^2
   \quad (s~~{\rm being~away~from}~{m_{J/\psi}}^2,..) ~.
\end{eqnarray}}
\item{In the intermediate region, there is a
characteristic interference 
between the LD contribution and
the SD contribution, 
which can be only derived from the detailed expression of
the distributions, Eq. (\ref{eq9}). }
\item{The value of the ratio does not  depend  much on 
whether the decaying particles are 
$B$ or $ \overline{B}$ 
at any invariant mass-squared region.}
\item{This ratio changes only a few $ \% $ when 
we change the input parameters, $m_t$ and $\mu$, within the range
shown in Table 1. The dependences on $m_t$ and $\mu$ 
are almost cancelled away in the ratio. 
Therefore, the uncertainties in this ratio due to the input 
parameters are much smaller than  the uncertainties in
the differential decay rate itself, 
as shown in Tables 2, 3 and 4 as well as 
in Figures \ref{fig:1} and \ref{fig:2}.}
\item{In Fig. \ref{fig:4} 
the range between the dot-dashed curve and the 
solid curve corresponds to the value of the ratio 
${1 \over \lambda^2} {dR(s)\over ds}$ allowed
from the present experimental result of 
$B_d - \overline{B}_d$ mixing.
Future experimental measurements 
on the ratio of the branching
fractions, 
${\cal B}(B \rightarrow X_d l^+ l^-)$ and
${\cal B}(B \rightarrow X_s l^+ l^-)$, 
can give much better alternative
for determination of 
$|{V_{td} \over V_{ts}}|$ without any hadronic
uncertainties, limited only by experimental statistics. }
\item{If $10^9$ $B \overline{B}$ pairs are produced,
the expected number of the events for
$B \rightarrow X_d l^+ l^-$ in the range of 
$4 {m_\mu}^2 < s < 6$ GeV (${m_\mu}=$ muon mass)
is about $100$ for  $(1-\rho)^2+\eta^2=0.59$, 
and is about $220$ for  $ (1-\rho)^2+\eta^2 =1.33$. 
Therefore, the statistical accuracy of  $(1-\rho)^2+\eta^2$ 
determined from this method is about $7 \% \sim 10 \%$
with the expected production of $10^9$ $B \overline{B}$ pairs.}  
\item{We have assumed in our numerical analysis 
the flavor $SU(3)$ symmetry with $m_d = m_s$.
We estimated the corrections due to $SU(3)$ breaking
by varying $m_d$ from 0.01 GeV up to $m_s = 0.2$ GeV.
And we find the ratio  ${dR \over ds} $ decreases within 
$0.2 \sim 0.3 \%$ for the range $1<s<9$ (GeV$^2$).
Therefore, we conclude the $SU(3)$ breaking effect does not 
affect the extraction of $|{V_{td} \over V_{ts}}|$ at all.}
\end{itemize}
\vskip 0.5cm
%

{\Large \bf Acknowledgements}\\

TM would like to thank G. Hiller, K. Ochi, T. Nasuno , and
Y. Kiyo for correspondence and assistance of
numerical computations.
The preliminary version of this work was given \cite{moro}
as a talk by TM at BCONF97, Hawaii, March 1997.
The work of CSK was supported
in part by the CTP, Seoul National University,
in part by the BSRI Program, Ministry of Education, 
Project No. BSRI-97-2425,
and in part by the KOSEF-DFG large collaboration project, 
Project No. 96-0702-01-01-2.
The work of TM was partially supported by 
Monbusho International Scientific Research
Program (No. $08044089$), and that of 
AIS and TM was supported also by Grant-in-Aid for 
Scientific Research on Priorty Areas (Physics of CP
violation) from the Ministry of Education and  Culture of Japan
.

\newpage
\begin{table}[h]
        \begin{center}
        \begin{tabular}{|l|l|}
        \hline
        \multicolumn{1}{|c|}{Parameter} &
                \multicolumn{1}{|c|}{Value}     \\
        \hline \hline
        $m_W$                   & $80.26$ (GeV) \\
        $m_Z$                   & $91.19$ (GeV) \\
        $\sin^2 \theta_W $      & $0.2325$ \\
        $m_s$                   & $0.2$ (GeV)   \\
        $m_d$                   & $0.01$ (GeV) \\
        $m_c$                   & $1.4$ (GeV) \\
        $m_b$                   & $4.8$ (GeV) \\
        $m_t$                   & $175 \pm 9$ (GeV)     \\
        $\mu$                 & $5^{+5.0}_{-2.5}$ (GeV) \\
$\Lambda_{QCD}^{(5)}$ & $0.214^{+0.066}_{-0.054}$ (GeV) \\
        $\alpha_{QED}^{-1}$     & 129           \\
        $\alpha_s (m_Z) $       & $0.117 \pm 0.005$ \\
        ${\cal B}_{sl}$         & $(10.4 \pm 0.4)$ \%   \\
        $\lambda_1$             & $-0.20$ (GeV$^2$) \\
        $\lambda_2$             & $+0.12$ (GeV$^2$) \\
        \hline
        \end{tabular}
        \end{center}
\caption{Values of the input parameters used in the numerical
          calculations of the decay rates. 
         Unless, otherwise specified,
          we use the central values.}
\label{parameters}
\end{table}
\begin{table}[h]
        \begin{center}
        \begin{tabular}{*{10}{|r}|}
\hline
                $\mu$ &
                $C_1$ &
                $C_2$ &
                $C_3$ &
                $C_4$ &
                $C_5$ &
                $C_6$ &
                $C_7^{\mbox{eff}}$ &
                $C_9^{NDR}$ &
                $C^{(0)}$ \\
\hline
                5 &
               -0.2404 &
                1.1031 &
                0.0107 &
                -0.0249&
                0.0072&
                -0.0302&
                -0.3110&
                4.1530 &
                0.3805 \\
\hline
               10 &
               -0.1606 &
                1.0642 &
                0.0068 &
                -0.0170&
                0.0051&
                -0.0194&
                -0.2768&
                3.7551&
                0.5816 \\
\hline
               2.5 &
               -0.3472 &
                1.1614 &
                0.0163 &
                -0.0348&
                 0.0096& 
                -0.0462&
                -0.3525&
                4.4128 &
                0.1163 \\
\hline
\end{tabular}
\end{center}
\caption{ $\mu$ (in GeV)
 dependence of the Wilson coefficients used in the numerical
      calculations. The values of $\Lambda_{QCD}$ and $m_t$ 
 are fixed at their central values.  }
\end{table}
\newpage
\begin{table}[h]
        \begin{center}  
        \begin{tabular}{*{10}{|r}|}
\hline
                ${\Lambda_{QCD}}^5$ &
                $C_1$ &
                $C_2$ & 
                $C_3$ & 
                $C_4$ & 
                $C_5$ & 
                $C_6$ & 
                $C_7^{\mbox{eff}}$&
                $C_9^{NDR}$ &
                $C^{(0)}$ \\
\hline
                0.214 &
               -0.2404 &
                1.1031 &
                0.0107 &
                -0.0249&
                0.0072&
                -0.0302&
                -0.3110&
                4.1530 &
                0.3805 \\
\hline
            0.280&
               -0.2579         &
                1.1122        &
                0.0116        &
               -0.0265         &
                0.0076      &
                -0.0327     &
               -0.3181         &
                4.2137        &
                0.3369  \\   
\hline
            0.160
 &
                -0.2242   &
                1.0949    &
                0.0099   &
                 -0.0233       &
                0.0068       &
                -0.0279       &
                -0.3043       &
                 4.0891      &
                 0.4212      \\
\hline
\end{tabular}
\end{center}
\caption{ ${\Lambda_{QCD}}^5$  dependence of 
          the Wilson coefficients 
          used in the numerical calculations. The values of
      $m_t$ and $\mu$ are fixed at their central values.  }
\end{table}
\begin{table}[h]
\begin{center}
        \begin{tabular}{|r|r|r|r|}
\hline
                $m_t$ &
                $C_7^{\mbox{eff}}$ &
                $C_9^{NDR}$ &
                $C_{10}$ \\
\hline
                $175$ &
                $-0.311   $ & 
                $4.153   $ &   
                $-4.546   $ \\ 
\hline
                $166$ &
                $-0.3066   $ &
                $4.0796   $ &
                $-4.1877   $ \\
\hline
                $184$ &
                $-0.3150   $ &
                $4.2238   $ &
                $-4.9156   $
\\
\hline
\end{tabular}
\end{center}
\caption{ $m_t$ dependence of 
          the Wilson coefficients used in the numerical
  calculations. The values of ${\Lambda_{QCD}}^5 $ and 
  $\mu$ are  fixed at their central values. }
\label{wilson}
\end{table}

{\Large \bf Appendix}

\appendix 
\section{Functions $g(z, \hat{s})$, 
                $g(0, \hat{s})$ and $\omega(\hat{s})$}

The functions $g(z, \hat{s})$ and 
$g(0, \hat{s})$ are given as
\begin{eqnarray}
g(z,\hat{s}) &=& -\frac{8}{9}\ln (\frac{m_b}{\mu})
 -\frac{8}{9} \ln z + \frac{8}{27} +\frac{4}{9}y
-\frac{2}{9}(2 + y) \sqrt{\vert 1-y \vert}\nonumber\\
&\times & 
\left[\Theta(1-y)(\ln\frac{1+\sqrt{1-y}}{1-\sqrt{1-y}} -i\pi )
+\Theta(y-1) 2 \arctan \frac{1}{\sqrt{y-1}} \right] ,
\end{eqnarray} 
with $y= 4 z^2 / \hat{s}$, and
\begin{equation}
g(0,\hat{s}) = \frac{8}{27}-\frac{8}{9}\ln (\frac{m_b}{\mu})
              -\frac{4}{9}\ln \hat{s} + \frac{4}{9}i\pi ~.
\end{equation}

The function $\omega(\hat{s})$ represents the
$O(\alpha_s)$ correction 
from the one-gluon exchange in the matrix element of
$ O_9$ \cite{JK89};
\begin{eqnarray}
\omega(\hat{s}) &=& 
-\frac{2}{9}\pi^2 -\frac{4}{3}{\mbox Li}_2(\hat{s})
-\frac{2}{3}
\ln \hat{s} \ln(1-\hat{s}) -
\frac{5+4\hat{s}}{3(1+2\hat{s})}\ln(1-\hat{s})\nonumber\\
&-& 
\frac{2\hat{s}(1+\hat{s})(1-2\hat{s})}
     {3(1-\hat{s})^2(1+2\hat{s})}
\ln \hat{s} + 
\frac{5 + 9\hat{s} -6\hat{s}^2}{6(1-\hat{s})(1+2 \hat{s})} ~.
\label{omegahats}
\end{eqnarray}

\section{Functions $f(\mc)$ and $\kappa(\mc)$}

The phase space function for 
$\Gamma (B \rightarrow X_c l \nu)$
in the lowest order ({\it i.e.}, parton model)
\begin{equation}
  f(\mc) = 1 - 8 \, \mc^2 + 8 \, 
        \mc^6 - \mc^8 - 24 \, \mc^4 \, \ln \mc ~.
        \label{eqn:fr}
\end{equation}
And  $\kappa(\mc)$ accounts for both the $O(\alpha_s)$ QCD 
correction to the semi-leptonic decay  
width and the leading order
$(1/m_b)^2$ power correction,
\begin{equation}
\kappa(\mc) = 1 - \frac{2 \alpha_s(m_b)}{3 \pi} g (\mc)
    + \frac{h(\mc)}{2 m_b^2} ~.
\end{equation}
The function $g(\mc)$ is given \cite{JK89},\cite{cskim} as
\begin{equation}
g(\mc) = (\pi^2-\frac{31}{4})(1-\mc)^2 + \frac{3}{2} ~, 
\end{equation}
and finally the function $h(\mc)$ is given \cite{georgi} as
\begin{equation}
h(\mc) = \lambda_1 + \frac{\lambda_2}{f(\mc)} 
   \left[ -9 +24 \mc^2
-72\mc^4 + 72\mc^6 -15\mc^8 -72 \mc^4 \ln \mc \right] ,
\label{eqn:ghr}
\end{equation}
where $\lambda_1~({\rm or}~-\mu_{\pi}^2)$ and 
$\lambda_2~({\rm or}~\mu_{_G}^2)$ denote the matrix element
of the higher derivative operators of heavy quark effective
theory, as defined in Eq. (\ref{hqetpar}).

\newpage

\end{document}